\documentstyle[epsf,12pt]{article}

\def\bc{\begin{center}}
\def\ec{\end{center}}
\def\be{\begin{equation}}
\def\ee{\end{equation}}
\def\myappendix{\par
 \setcounter{section}{0}
 \setcounter{subsection}{0}
 \setcounter{equation}{0}
 \setcounter{table}{0}
 \def\appendixname{Appendix}
 \def\appesection{\setcounter{equation}{0}\section}
 \def\@thesection{\Alph{section}}
 \def\thesection{\appendixname\hskip 1.10ex\Alph{section}}
 \def\thesubsection{\@thesection.\arabic{subsection}}
 \def\theequation{\@thesection.\arabic{equation}}
 \def\thetable{\@thesection.\arabic{table}}}

\newcommand{\beq}{\begin{equation}}
\newcommand{\eeq}{\end{equation}}
\newcommand{\beqn}{\begin{eqnarray}}
\newcommand{\eeqn}{\end{eqnarray}}

\def\vdir{v\kern-7.8pt\Big{/}}
\def\pdir{p\kern-7.8pt\Big{/}}

\newcommand{\mfrac}[2]{\frac{\textstyle #1}{\textstyle #2}}
\begin{document}
\pagestyle{empty} 
\vspace{-0.6in}
\begin{flushright}
ROME prep. 96/1153 \\
FTUV 96/26 - IFIC 96/30
\end{flushright}
\vskip 0.2 cm
\centerline{\LARGE{\bf{$B$--$\bar B$ Mixing in the HQET}}}
\vskip 1.4cm
\centerline{\bf{V. Gim\'enez$^{1}$,
G. Martinelli$^{2}$}}
\centerline{$^1$ Dep. de Fisica Teorica and IFIC, Univ. de Valencia,}
\centerline{Dr. Moliner 50, E-46100, Burjassot, Valencia, Spain.} 
\centerline{$^2$ Dip. di Fisica, Univ. ``La Sapienza"  and}
\centerline{INFN, Sezione di Roma, P.le A. Moro, I-00185 Rome, Italy.}
\abstract{
We present a high statistics, quenched lattice calculation of the 
$B$-parameters  $B_{B_d}$ and $B_{B_s}$, computed  
at lowest order in the HQET.  The  results were
obtained using a sample of $600$ quenched gauge field configurations,
generated by Monte Carlo simulation 
at $\beta=6.0$   on a $24^{3}\times 40$ lattice.
For the light quarks  the SW-Clover action was used; the propagator
of the lattice HQET was also tree-level improved.
Our best estimate of the
renormalization scale independent $B$-parameter is 
$\hat{B}_{B_d } = 1.03\, \pm\, 0.06\pm 0.18$.  $\hat{B}_{B_d }$
has been   obtained by using ``boosted'' perturbation theory 
to calculate the renormalization constants  which relate the matrix elements
of  the lattice operators  to the corresponding amplitudes in the continuum. 
Due to the large statistics, the errors in the extraction
of the  matrix elements of the relevant bare operators are
rather small.
The main systematic error, corresponding to $\pm 0.18$ in the 
above result,  comes  from the uncertainty in the evaluation
of the renormalization constants, for   which the one-loop corrections
are rather large. The non-perturbative evaluation of these
constants will help to reduce the final error.
We also obtain $\hat{B}_{B_s }/\hat{B}_{B_d }=1.01 \pm 0.01$ and
$f^2_{B_s}\hat{B}_{B_s }/f^2_{B_d}\hat{B}_{B_d }=1.38 \pm 0.07$.
}
\vfill\eject
\pagestyle{empty}\clearpage
\setcounter{page}{1}
\pagestyle{plain}
\newpage 
\pagestyle{plain} \setcounter{page}{1}
 
\section{Introduction} \label{intro}
The study of  $B_{d}^{0}$--$\bar{B}_{d}^{0}$ and
 $B_{s}^{0}$--$\bar{B}_{s}^{0}$ mixings may narrow the
constraints on the elements of the Cabibbo-Kobayashi-Maskawa (CKM)
matrix, which in turn serve to test the Standard Model, and to detect
possible new physics beyond it \cite{analy}. 
 The results of the  phenomenological analyses which involve
$B^0$--$\bar B^0$ mixings~\footnote{ 
 The subcripts  $d$ or $s$
referring to the light quark flavour will be written
explicitely only when necessary for the
discussion.}  strongly depend on two
non-perturbative inputs: the value of the decay constant of the $B$-meson,
$f_{B}$, and the value of the renormalization
group invariant $B$-parameter, $\hat B_{B}$. These
parameters enter in the theoretical predictions for  
$x_{d}$ and $x_{s}$, to be defined below,
 through the mixing parameter 
 $\xi_{B}\equiv f_{B}\, \sqrt{\hat B_{B}}$, $\xi^2_{B}$ corresponding, up
to trivial factors, to the matrix elements of the relevant
$\Delta B=2$ operators. 
The $B$-meson decay constant has not been
measured yet. However, several theoretical evaluations of $f_{B}$ using 
numerical simulations of  lattice QCD~\cite{blattice} and
 QCD sum rules~\cite{qcdsr}
have been published to date.  The $B$-parameter, which is equal to one
in the vacuum saturation approximation, has also been estimated using
the same non-perturbative methods \cite{blattice,qcdsr}.
\par In this paper a high statistics lattice calculation of the
$B$-parameter, at lowest order in the HQET, is presented. 
The relevant matrix elements
have been computed on  a statistical sample of $600$ gauge field configurations.
This has allowed us  to study two- and three-point correlation
functions at large time-distances with relatively
small statistical errors, thus   reducing the systematic
errors due to the contamination of higher excited states.
For  the  bare four-fermion operators,
it was possible to isolate  a good signal
by using cube and double-cube smeared axial currents as the
interpolating fields  for the $B$-mesons, and to extract their matrix
elements  with small statistical errors.
The major source of systematic error is presently given by
the uncertainties in the evaluation of the renormalization
constants of the different operators, which are known
only at first order in perturbation theory. Indeed,  as explained
below, 
by renormalizing the operators in two different ways, which are
equivalent up to $O(\alpha_s^2)$, we obtain  either
\beq  \hat{B}_{B_{d}}  =  1.21 \pm 0.08 \, \mbox{ Method 1} \label{twores1}
\eeq
or
\beq \hat{B}_{B_{d}}  =   0.86 \pm 0.06 \, \mbox{ Method 2} \, .
\label{twores2} \eeq
By combining eqs.~(\ref{twores1}) and (\ref{twores2}), we derive our
final result for this quantity 
\beq  \hat{B}_{B_{d}}  =  1.03 \pm 0.06 \pm 0.18 \label{final} \,  ,
\eeq
where the last error is an estimate of the uncertainty due
to the contribution of  higher-order terms to the renormalization
constants. \par 
The ratio of $B$-parameters $\hat{B}_{B_{s}}
/\hat{B}_{B_{d}}$, which may be obtained by studying the
dependence of the $B$-parameter on the light quark mass, has also
been considered.
In this ratio,  most of the uncertainties due to higher-order terms
cancel and we quote
\be \frac{ \hat{B}_{B_{s}}}{\hat{B}_{B_{d}}}=1.01 \pm 0.01 \label{ratio}
\, .\ee
By using $f_{B_s}/f_{B_d}=1.17 \pm 0.03 $, as derived from the study of the
two-point correlation functions on the same set of configurations,
and $M_{B_s}/M_{B_d}=1.017$ we then obtain
\beq  r_{sd}=\frac{f^2_{B_s}M_{B_s}^2 \hat{B}_{B_{s}}}{f^2_{B_d}
M_{B_d}^2 \hat{B}_{B_{d}}}
= 1.43 \pm 0.07 \, .\eeq
$r_{sd}$ can also be obtained by taking directly the ratio of the 
matrix elements of the $\Delta B=2$ operator, computed at the appropriate
value of the quark masses. In  this case we get
\beq  r_{sd}=\frac{f^2_{B_s}M_{B_s}^2 \hat{B}_{B_{s}}}
{f^2_{B_d}M_{B_d}^2\hat{B}_{B_{d}}}
= 1.35 \pm 0.05 \, .\eeq
smaller than the result $r_{sd}=1.54(13)$ 
obtained using propagating  Wilson quarks instead
 of the HQET~\cite{rsd}~\footnote{ In 
ref.~\cite{rsd}, they also presented a result which includes the error
due to the extrapolation to $a \to 0$. Since our
result has been obtained only at $a^{-1}\sim 2$ GeV,
we compare it with the result of ref.~\cite{rsd} obtained by
fitting $r_{sd}$ to a constant in $a$.}. The experimental values of
$M_{B_d}$ and $M_{B_s}$ were taken from ref.~\cite{pdg}.
\par  Only  a non-perturbative calculation of 
the renormalization constants,  which relate the relevant lattice operators
to the continuum one, can reduce the present uncertainty in the 
prediction of the $B$-parameters from the HQET.  This computation is in 
progress and the results will be published elsewhere.

\par
The plan of the paper is the following. In sec.~\ref{bbdef}
the relevant formulae for the definition of $\hat B_{B}$ are given; in 
sec.~\ref{proc}
the procedure followed to compute  the $B$-parameter is presented; 
in sec. \ref{numerical} the
numerical calculations of the relevant matrix elements
are  described and the main
results of this study are discussed; 
in the conclusions an outlook of
possible developments of the calculations of this paper is given. 

\section{Master formulae for  $B^{0}$--$\bar B^{0}$ mixings} 
\label{bbdef}

$B^{0}$--$\bar B^{0}$ mixings are  usually expressed as
\be
x_{d,s}\equiv\mfrac{\left(\triangle M\right)_{B_{d,s}}}{\Gamma_{B_{d,s}}}
\ ,\ee
where $\left(\triangle M\right)_{B_{d,s}}$ is the mass difference of the
$B^{0}_{d,s}$--$\bar B^{0}_{d,s}$ system and 
$\Gamma_{B_{d,s}}=1/\tau_{B_{d,s}}$ is the average total width.
Using the effective $\Delta B=2$ Hamiltonian discussed in ref.~\cite{buras}
 one finds
\be x_{d,s}\, =\, \tau_{B_{d,s}}\, \mfrac{G_{F}^{2}}{6\pi^{2}}\,\eta_{B}\,
M_{B_{d,s}}\, \xi^{2}_{B_{d,s}}\, M_{W}^{2}\, S_{0}(x_{t})\, \vert
V_{t(d,s)}\vert^{2} \ ,
\label{buri} \ee
where $\eta_{B} \sim 0.55$, $S_{0}(x_t)$ is a  function of the ratio
$x_{t}=m_{t}^{2}/M_{W}^{2}$, $m_{t}$ and $M_W$ 
are  the top quark and $W$ masses respectively, $G_F$ the Fermi constant,
$V_{t(d,s)}$ the Cabibbo-Kobayashi-Maskawa matrix element  and
$\xi_{B_{d,s}}=f_{B_{d,s}}\, \sqrt{\hat{B}_{B_{d,s}}}$. 
\par The renormalization
group invariant parameter $\hat{B}_{B_{d,s}}$ is defined by
\be  \hat{B}_{B_{d,s}}\, =\, \alpha_{s}(\mu^2)^{-6/23}\, \left[\, 1\,+\,
\mfrac{\alpha_{s}(\mu^2)}{4\, \pi}\, J_{5}\,\right]\,\, B_{B_{d,s}}(\mu)
\label{bbhat}\ , \ee
where 
\be
B_{B_{d,s}}(\mu)\, =\, \mfrac{\langle\bar B^{0}_{d,s}|\hat 
O_{L}(\mu)|B^{0}_{d,s}\rangle}{
\mfrac{8}{3}\, f_{B_{d,s}}^{2}\, M_{B_{d,s}}^{2}}\  \ee
$\mu$ is the operator renormalization scale and  $f_{B_{d,s}}$ is 
the $B_{d,s}$-meson  decay constant 
\be
\langle 0| \bar{b}(0)\, \gamma_{\mu}\, \gamma_{5}\, q(0) | B(\vec{p})\rangle\, =\, i\, p_{\mu}\, f_{B}
\ . \ee
 $\hat O_{L}$ is the renormalized four--quark operator 
\be
\hat O_{L}\, =\, \left(\, \bar{b}(x)\, \gamma^{\mu}\, (1 -\gamma_{5})\, q(x)\, \right) 
\, \left(\, \bar{b}(x)\, \gamma_{\mu}\, (1 -\gamma_{5})\, q(x)\, \right) 
\ , \ee
and $b(x)$ and $q(x)$ are  the heavy and light quark fields respectively.

The factor $J_{5}$ can be written in terms of the 
one- and two-loop anomalous dimensions, $\gamma^{(0)}$ and 
$\gamma^{(1)}$, of the operator $\hat O_{L}$; if $\hat 
O_L(\mu)$ is renormalized
in the $\overline{MS}$-scheme one gets \cite{buras}
\begin{eqnarray}
\gamma^{(0)} &=&
4\ , \;\;\;\;\;\;\;\;\;
\gamma^{(1)}=-7\,+\,\mfrac{4}{9}\, n_{f}\ , \;\;\;\;\;\;\;\;\;
\beta_{0} = 11\,-\, \mfrac{2}{3}\, n_{f}\ ,\nonumber\\
 \beta_{1}&=&102\,
-\, \mfrac{38}{3}\, n_{f}\ , \;\;\;\;\;\;\;\;\;
J_{n_{f}} = \mfrac{\gamma^{(0)}\,\beta_{1}}{2\,\beta_{0}^{2}}\, -\,
\mfrac{\gamma^{(1)}}{2\,\beta_{0}} \ .
\end{eqnarray}
$\beta_{0,1}$ are the one- and two-loop coefficients of the $\beta$
function and $n_{f}$ the number of active light quark flavours. 
$n_{f}=5$ for $\mu \ge m_b$. 
Notice that, strictly speaking, $\hat{B}_{B}$ is renormalization 
scale independent only up to the next-to-leading order.  
\par
Since $m_b$ is larger than the current values of the lattice cut-off, 
in order to predict physical quantities relevant in $B$-physics
a possibility is to use the lattice version of the
HQET \cite{eichten}, which allows
a systematic expansion in inverse powers of the heavy quark mass.
It this case, to obtain the physical matrix elements of the 
effective Hamiltonian, one has  to match
appropriate lattice bare operators to the operator $\hat O_L(\mu)$,
which is renormalized in the continuum ``full'' theory. The
matching procedure is conventionally splitted in two separate 
steps~\footnote{
As explained in the first of refs.~\cite{analy} this is, however, unecessary.}:\par
i) the matching of the continuum HQET to the full theory; \par 
ii) the matching of the lattice HQET to its  continuum counterpart.\par 
The two-loop Next-to-Leading Order
(NLO) anomalous dimension  necessary for step i) 
has been computed  using the  $\overline{MS}$ dimensional regularization 
in ref.~\cite{vice}.  As noticed in ref.~\cite{bu2}, the calculation was,
however, not complete because some elements of the one-loop
mixing matrix were missing. This calculation 
has been recently completed in refs.~\cite{cfg,buch}.   
The final result can be written as 
\be \hat O_L(m_b)= C_L(\mu^2) \tilde{O}_L(\mu) + C_S(\mu^2) \tilde{O}_S(\mu)
\ , \label{uno} \ee
where $\hat O_L(m_b)$ is the operator in the full theory renormalized 
in the $\overline{MS}$-scheme at the
a scale equal to $b$-quark mass 
($\hat O_L(m_b)=\hat O_L(\mu=m_b)$) and 
\begin{eqnarray}
\tilde{O}_{L} &=& 2\, \Bigl(\, \bar{h}(x)\, \gamma^{\mu}\,
 (1 -\gamma_{5})\, q(x)\, \Bigr) 
\, \Bigl(\, \bar{h}^{(-)}(x)\, \gamma_{\mu}\, (1 -\gamma_{5})\, q(x)\,
\Bigr)\  \nonumber\\ 
\tilde{O}_{S} &=& 2\, \Bigl(\, \bar{h}(x)\, (1 -\gamma_{5})\, q(x)\, \Bigr) 
\, \Bigl(\, \bar{h}^{(-)}(x)\, (1 -\gamma_{5})\, q(x)\, \Bigr) \ 
\label{ols} \end{eqnarray}
are the operators of the effective theory, renormalized at the scale $\mu$.  
The fields $\bar h$ and $\bar h^{(-)}$ create a heavy-quark or annihilate
 a heavy anti-quark state respectively.
The coefficient functions are given by \cite{cfg,buch}
\begin{eqnarray} \label{match1}
C_L(\mu^2) &=& 
\left(\mfrac{\alpha_{s}(m^2_{b})}{\alpha_{s}(\mu^2)}\right)^{d_1}\, 
\left(1\,+\, \mfrac{\alpha_{s}(\mu^2)-\alpha_{s}(m^2_{b})}{4\pi}\,
 J\, \right) C_L(m_b^2) \nonumber\\
&+& \left[\left(\mfrac{\alpha_{s}(m^2_{b})}{\alpha_{s}(\mu^2)}\right)^{d_2}
-\left(\mfrac{\alpha_{s}(m^2_{b})}{\alpha_{s}(\mu^2)}\right)^{d_1} \right]
\frac{\hat \gamma^{(0)}_{21}}{\hat \gamma^{(0)}_{22}-\hat\gamma^{(0)}_{11}}
C_S(m_b^2) \ ,  \nonumber \\
C_S(\mu^2) &=& 
\left[\mfrac{\alpha_{s}(m^2_{b})}{\alpha_{s}(\mu^2)}\right]^{d_2}\, 
C_S(m_b^2) \ , \label{coe}
\end{eqnarray}
where, in n\"aive dimensional regularization,  
\beqn C_L(m_b^2) = 1 - 14 \frac{\alpha_s(m_b^2)}{4 \pi}\ ,
 \;\;\;\;\;\;\;\;\;
  C_S(m_b^2) = -8  \frac{\alpha_s(m_b^2)}{4 \pi}\ , \label{cou} \eeqn
with
\beq d_i= \frac{\hat \gamma^{(0)}_{ii}}{2 \beta_0}
\;\;\;\;\;\;\;\;\; J =\beta_1 \frac{d_1}{\beta_0}-
 \frac{\hat \gamma^{(1)}_{11}}{2 \beta_0} \ .  \eeq
The elements of the one-loop anomalous mixing matrix
are given by
\beqn \hat  \gamma^{(0)}_{11} &=& -8 \;\;\;\;\;\;\;\;\;
\hat \gamma^{(0)}_{12} = 0 \ ,  \;\;\;\;\;\;\;\;\;
\hat \gamma^{(0)}_{21} = \frac{4}{3}\;\;\;\;\;\;\;\;\; 
\hat \gamma^{(0)}_{22}=-\frac{8}{3} \eeqn
and
\beq
\hat{\gamma}^{(1)}_{11}=-\mfrac{808}{9}\,-\,\mfrac{52}{27}\,\pi^{2}\,+\,
\mfrac{64}{9}\, n_{f} \ .\eeq
\par We  now consider step ii), 
i.e. the matching of the lattice to the continuum HQET. 
Owing to the breaking of 
chiral symmetry induced by the Wilson term, the discretized version of
$\tilde{O}_{L}$,  mixes with two new 
lattice operators as follows \cite{oneloop}--\cite{borre2}
\begin{eqnarray} 
\tilde{O}_{L}(\mu) &=&  \left( 1\,+\, \mfrac{\alpha^{L}_{s}}{4 \pi}
[4\, \ln(a^{2} \mu^{2})\, +\, D_{L}\,] \right)\, O_{L}(a)\nonumber\\
&+& \mfrac{\alpha^{L}_{s}}{4 \pi}\, D_{R}\, O_{R}(a)\, +\,
\mfrac{\alpha^{L}_{s}}{4 \pi}\, D_{N}\, O_{N}(a) \label{ma2} \\
\tilde{O}_{S}(\mu) &=&  \left( 1\,+\, \mfrac{\alpha^{L}_{s}}{4 \pi}
[\frac{4}{3}\, \ln(a^{2} \mu^{2})\, +\, D_{S}\,] \right)\, O_{S}(a) 
+ \dots\label{mas}
\end{eqnarray}
where $O_i(a)$ denote a bare lattice operator;
\begin{eqnarray} \label{match2}
\mfrac{1}{2}\, O_{R}(a) &=& \left(\, \bar{h}(x)\, \gamma^{\mu}\, (1 +\gamma_{5})\, q(x)\, \right) 
\, \left(\, \bar{h}^{(-)}(x)\, \gamma_{\mu}\, (1 +\gamma_{5})\, q(x)\,
\right)\ ; \nonumber\\
\mfrac{1}{2}\, O_{N}(a) &=& \left(\, \bar{h}(x)\, \gamma^{\mu}\, (1 -\gamma_{5})\, q(x)\, \right) 
\, \left(\, \bar{h}^{(-)}(x)\, \gamma_{\mu}\, (1 +\gamma_{5})\, q(x)\,
\right)\nonumber\\
&+&\left(\, \bar{h}(x)\, \gamma^{\mu}\, (1 +\gamma_{5})\, q(x)\, \right) 
\, \left(\, \bar{h}^{(-)}(x)\, \gamma_{\mu}\, (1 -\gamma_{5})\, q(x)\,
\right)\nonumber\\ 
&+& 2\, \left(\, \bar{h}(x)\, (1 -\gamma_{5})\, q(x)\, \right) 
\, \left(\, \bar{h}^{(-)}(x)\, (1 +\gamma_{5})\, q(x)\,
\right)\nonumber\\ 
&+& 2\, \left(\, \bar{h}(x)\, (1 +\gamma_{5})\, q(x)\, \right) 
\, \left(\, \bar{h}^{(-)}(x)\, (1 -\gamma_{5})\, q(x)\,
\right) \ ; \label{ovari}
\end{eqnarray}
$\alpha^{L}_{s}$ is the lattice  coupling constant and
$D_{L}, \dots \ , D_R$  are constants which have been computed  
in one-loop lattice perturbation theory. The leading logarithmic correction
to $\tilde{O}_S$ in eq.~(\ref{mas}) is taken into account by renormalizing the
operators at a scale $\mu=1/a$; the term $D_S$ and the $\dots$ represent
finite terms of $O(\alpha_s)$. Since $\tilde{O}_S$ enters only  at the NLO, they do
not need to be computed, cf. eqs.~(\ref{coe}) and (\ref{cou}).
\noindent
Two  remarks are in order here: 
\begin{itemize}
\item The value of the constants $D_i$ depend on the light quark action
used in the numerical simulation. In our
case the tree-level improved  SW-Clover  
action  \cite{clover} was used  and the 
operators were improved by rotating the light-quark 
propagators \cite{impimp}.
\item In the calculation of the coefficients $D_i$, 
the heavy-quark field renormalization constant computed 
in   ref.~\cite{pene1}--\cite{fitfit} was used.
This is the definition which is consistent with improvement \cite{fitfit}
and  with the method used to extract the matrix elements
of the operators $O_i(a)$, see also the discussion in 
refs.~\cite{l12,ukqcdbb}.
\end{itemize}
The details of the numerical calculation of the coefficients 
$D_i$ can be found in ref.~\cite{borre2}. 
Here, we only quote the final results:
\be
D_{L}=-22.5\;\;\;\;\;\;\;\;\;D_{R}=-5.4\;\;\;\;\;\;\;\;\;D_{N}=-14.0 \label{dconst} 
\ee

Lattice perturbation theory behaves rather badly due to the presence
of tadpole-like diagrams which are typical of the lattice case~\footnote{
The inaccuracy
of lattice perturbation theory has
 been  demonstrated  explicitely in several cases where a non-perturbative
determination of the renormalization constants was possible
\cite{msv0,lm}.}.
For logarithmically divergent operators,
two remedies to this problem have been proposed in the recent
years, namely
 boosted perturbation theory \cite{lm} or  the non-perturbative 
renormalization of the operators on quark and gluon states,  in a fixed gauge
\cite{mpstv}. Only the first method can be used,
since the non-perturbative calculation of the renormalization
constants relevant in the present study has not been performed yet. 
\par In eq.~(\ref{ma2}),  for 
comparison with the recent calculation of the UKQCD
collaboration \cite{ukqcdbb}, 
$\alpha^{L}_{s}=6/4\pi\beta u_{0}^{4}$ has been used.
$u_{0}$ is a measure of
the average link variable,  $u_{0}=(8 K_{c})^{-1}$,
with $K_{c}$ the value of the Wilson hopping parameter $K$
at which the pion mass vanishes. In the SW-Clover case, the numerical 
value of the ``boosted'' coupling given above is very close to the 
boosted coupling defined in terms of the elementary plaquette
(see
ref.~\cite{lm} for the different definitions).  We have also used
$\alpha_s^L=
\alpha_V(q^*)$, where $\alpha_V(q^*)$ was also introduced in ref.~\cite{lm},
and $q^*$ is an appropriate scale, which can be extracted in one-loop 
perturbation theory.  For $q^*$, 
the value computed for the heavy-light
axial-vector  current for  the Wilson action in 
ref.~\cite{q*h} was used, $q^* a=2.18$, with $a^{-1}=2$ GeV,
since the appropriate  result for the present case
is not known.  \par 
Putting the  corrections of steps i) and ii)  together, 
see eqs.~(\ref{coe}), (\ref{ma2}), (\ref{mas})  and (\ref{dconst}),
the renormalized
operator $\hat O_{L}(m_b)$ can be written in
terms of the bare lattice operators $O_i(a)$ as follows:
\beq \label{zpt1}
\hat O_{L}(m_{b})\,=\, Z_{O_{L}}\, O_{L}(a)\, +\,
Z_{O_{R}}\, O_{R}(a)\, +\, Z_{O_{N}}\, O_{N}(a)\, +\,
Z_{O_{S}}\, O_{S}(a) \ .
\eeq
\par The renormalization constants $Z_{O_i}$ are obtained
from products of the coefficients in eqs.(\ref{coe}),
(\ref{ma2}) and (\ref{mas}), 
which are computed  independently  and
 contain terms of $O(\alpha_s)$. Thus there are two different ways
in which one can organize the final results, 
i.e. by including ($M_1$) or excluding ($M_2$)
next-to-next-to-leading terms of $O(\alpha_s^2)$ (without logarithms).
Since the corrections are large, different choices will result
in $B$-parameters which differ by about $28 \%$. With the present
state of art this is an intrinsic systematic error that can be reduced only
by incresing the order of the perturbative calculation in the continuum
and applying the non-perturbative renormalization to the
lattice operators.   \par  In eq.~(\ref{coe}),
the $Z_{O_i}$ have been evaluated by using $m_{b}=5$ GeV, 
$\Lambda^{n_f=4}_{QCD}=200$ MeV, $\mu=2 $ GeV $ \sim 1/a$ and $n_f=4$.
At the NLO this choice of the parameters gives $\alpha_s(m^2_b)=0.1842$, which
is also the value used in eq.~(\ref{bbhat}), since the matching
between the full theory and the effective one
is made  at $\mu=m_b$.

In eq.~(\ref{ma2}) 
three possible values for $\alpha^L_s$ were used, and
the mixing coefficients $Z_{O_i}$ were computed 
with both options $M_1$ and $M_2$.
The values of the $Z_{O_i}$, together with the corresponding values
of $\alpha_s^L$ are given in table \ref{tab:zoi}. In the same
table we also give the renormalization constant of the heavy-light
axial current, computed in ref.~\cite{gim1}
at the NLO,  which is needed for the calculation of the 
$B$-parameter, see secs.~\ref{proc} and \ref{numerical}. 
\begin{table} \centering 
\begin{tabular}{||c|c|c|c|c|c|c||}
\hline
\hline
\multicolumn{2}{||c|}{Options}&\multicolumn{1}{c|}{$Z_{O_L}$}&
\multicolumn{1}{c|}{$Z_{O_R}$}&\multicolumn{1}{c|}{$Z_{O_N}$}&
\multicolumn{1}{c|}{$Z_{O_S}$}&\multicolumn{1}{c||}{$Z_{A}$} \\
\hline
\hline
$\alpha_s^L$  Standard & $M_1$ &  .7949& -.0317& -.0822& -.1229&  .9035\\ \cline{2-7}
  .0796& $M_2$ &  .7651& -.0394& -.1021& -.1229&  .8989\\ \hline
$\alpha_s^L$  Boosted $u_0$ & $M_1$ &  .6850& -.0581& -.1506& -.1229&  .7932\\ \cline{2-7}
  .1458& $M_2$ &  .6286& -.0722& -.1871& -.1229&  .7847\\ \hline
$\alpha_s^L$  Boosted V & $M_1$ &  .6283& -.0717& -.1859& -.1229&  .7363\\ \cline{2-7}
  .1800& $M_2$ &  .5581& -.0891& -.2310& -.1229&  .7257\\ \hline
\hline
\hline
\end{tabular}
\caption{\it{Renormalization constants for different choices of the
lattice coupling constants and options, see the text.}}
\label{tab:zoi}
\end{table}
\par Two observations are necessary at this
point.
\begin{itemize} \item In ref.~\cite{ukqcdbb} they have used the option $M_2$,
without, however, including the next-to-leading corrections of step i).
Moreover,  they took for
 $\alpha_s$ the value given by its leading logarithmic expression.
This is not consistent at the next-to-leading order and 
introduces a further (and easily avoidable) systematic effect
of $\sim 10 \%$ in the final result. 
\item Since, in absence of a better control of perturbation theory,
the uncertainties are so large, it is not worth at this point
to vary the value of $\Lambda_{QCD}$, or to worry about ``quenching",
 or
other more subtle questions.
\end{itemize}
\section{Computation of $B_{B}$ on the lattice.} 
\label{proc}

In order to obtain $B_{B}$, we compute the following two-
and three-point correlation  functions
\be
C^{RR'}(t)\, =\, \sum_{\vec{x}}\, \langle 0 | A^{R}_{0}(\vec{x},t)\,
 A^{R'\, \dag}_{0}(\vec{0},0) | 0 \rangle
\ ,  \label{cdef2} \ee
\be 
C^{RR'}_{O_{i}}(t_1,t_2)\, =\, 
\sum_{\vec{x}_1,\vec{x}_2}\, \langle 0 | A^{R}_{0}(\vec{x}_1,t_1)\,
O_{i}(\vec{0},0)\, A^{R'\, \dag}_{0}(\vec{x}_2,t_2) | 0 \rangle
\ , \label{cdef3} \ee
where $i=$L,R,N,S denotes  one of the operators on the r.h.s. of
eq.~(\ref{zpt1}). The labels 
 $R,R'=$L,S,D correspond to   different 
interpolation operators used for the $B$-mesons
\begin{eqnarray}
A^{L}_{\mu}(x) &=& \bar{h}(x)\,\gamma_{\mu}\, \gamma_{5}\, q(x)\ ,\nonumber\\
A^{S}_{\mu}(x) &=& \sum_{i} \bar{h}(x_{i})\,\gamma_{\mu}\, \gamma_{5}\, q(x)
\ , \\
A^{D}_{\mu}(x) &=& \sum_{i,j} \bar{h}(x_i)\,\gamma_{\mu}\, \gamma_{5}\, q(x_j)
\ , \label{sources} \end{eqnarray}
i.e.  the local (L), the cube-smeared (S) and
the double cube-smeared (D) axial currents
\cite{alltu}. Correlators involving smeared sources were computed
in the Coulomb gauge. The last two operators 
were
proven to be effective for isolating the lightest $B$-meson state.
\par
At large time distances the correlation functions above behave as
\be 
C^{RR'}(t)\, \longrightarrow\,  
Z^{R}\, Z^{R'}\,
e^{-\triangle E\, t}\ , \label{sinpar}
\ee
and
\be
C^{RR'}_{O_{i}}(t_1,t_2)\, \longrightarrow\, 
Z^{R}\, Z^{R'}\, \mfrac{\langle
\bar B^{0}|O_{i}(a )|B^{0}\rangle}{2\, M_{B}}
e^{-\triangle E\, (-t_1+t_2)}
\ .\label{sinpar2}\ee
 In the above equations
\be
Z^{R}\,=\, \mfrac{1}{\sqrt{2\, M_{B}}}\, 
\langle 0 |  A^{R}_{0}(\vec{0},0) | B \rangle \ ,
\ee
and $\triangle E$ is the binding energy of the B-meson.
To extract the matrix elements,  one takes the ratio
\be R^{RR'}_{O_{i}}(t_1,t_2)\, =\, \mfrac{C^{RR'}_{O_{i}}(-t_1,t_2)}{
\mfrac{8}{3}\, C^{RL}(-t_1)\, C^{R'L}(t_2)}
\ , \label{ratio1} \ee
which at large time distances behaves as
\be
R^{RR'}_{O_{i}}(t_1,t_2)\, \longrightarrow \, B^{RR'}_{O_{i}}\, \equiv\,
\mfrac{\langle \bar B^{0}|O_{i}(a)|B^{0}\rangle}
{\mfrac{8}{3}\, M_{B}^{2}\, f_{B}^{2}\,
Z_{A}^{-2}} \ , 
\ee
where $Z_{A}$ is the renormalization constant  for the axial current.
\par The 
physical value of the $B$-parameter is then given by
\be
B^{RR'}_{B}(m_{b})\, =\, 
\sum_{i=L,R,N,S}\, Z_{O_{i}}\, Z_{A}^{-2}\, B^{RR'}_{O_{i}}
\ . \ee
We  call this method the ratio method for determining the $B$-parameter.
One may also  consider the ratio
\be
R^{RR'}_{B}(t_1,t_2)\, =\, \sum_{i=L,R,N,S}\, 
Z_{O_{i}}\, \mfrac{C^{RR'}_{O_{i}}(-t_1,t_2)}{
\mfrac{8}{3}\, C^{RL}(-t_1)\, C^{R'L}(t_2)}
\ , \label{ratio2} \ee
which at large time distances behaves as
\be
R^{RR'}_{B}(t_1,t_2)\, \longrightarrow \, Z_{A}^{2}\, B_{B}^{RR'}(m_{b})\, 
\equiv\,
\mfrac{\langle
\bar B^{0}|\hat O_{L}(m_{b})|B^{0}\rangle}
{\mfrac{8}{3}\, M_{B}^{2}\, f_{B}^{2}\,
Z_{A}^{-2}} \ .  \ee
We call this method the combined-ratio method.

Both the ratio and the combined-ratio method should give the same 
value for  $B_{B}$. This is a  check of our numerical results.

\section{Numerical results.} 
\label{numerical}
As explained in the previous section, the determination of $B_{B}$
requires the computation of the two- and three-point correlation functions 
(\ref{cdef2}) and (\ref{cdef3}).
The SW-Clover fermion action \cite{clover} for the light quarks
was used, in the quenched approximation. The 
tree-level improved \cite{borre1,borre2} 
propagators of the heavy quarks were
computed in the static limit.
Our results are based on a set of $600$ gauge field configurations, computed
on a lattice of size $24^{3}\times 40$ at $\beta=6.0$.
The calculations were performed at three values of
the masses of the light quarks, corresponding to 
$K=0.1425,0.1432$ and $0.1440$.
This allows to
extrapolate the results  to the chiral limit.
All the errors have been computed  with the jacknife method by
decimating 30 configurations at a time.

The procedure to measure $B_{B}$ is standard.
At fixed $t_{1}$ ($t_{2}$), we study the behaviour of the ratios
(\ref{ratio1}) and (\ref{ratio2}) as a function of $t_{2}$ ($t_{1}$),
searching for a plateau in $t_{2}$ ($t_{1}$). $B_{B}(m_{b})$ is defined by 
the weighted average of the data points in the plateau region, if this exits. 
We will take as our best determination of $B_{B}(m_{b})$, the value evaluated 
in a time interval where the ratios appear to be independent of both 
$t_{1}$ and $t_{2}$. Notice that, contrary to the
UKQCD collaboration,  we never compute the matrix elements
of parity-odd terms of the different operators in eqs.~(\ref{ols})
and (\ref{ovari}), because they are  zero by parity. For this reason
the matrix element of $O_L$ and $O_R$ are equal and in the following
only the results for $O_L$ will be given.

\begin{figure}
\vspace{9pt}
\begin{center}\setlength{\unitlength}{1mm}
\begin{picture}(160,80)
\put(30,-20){\epsfbox{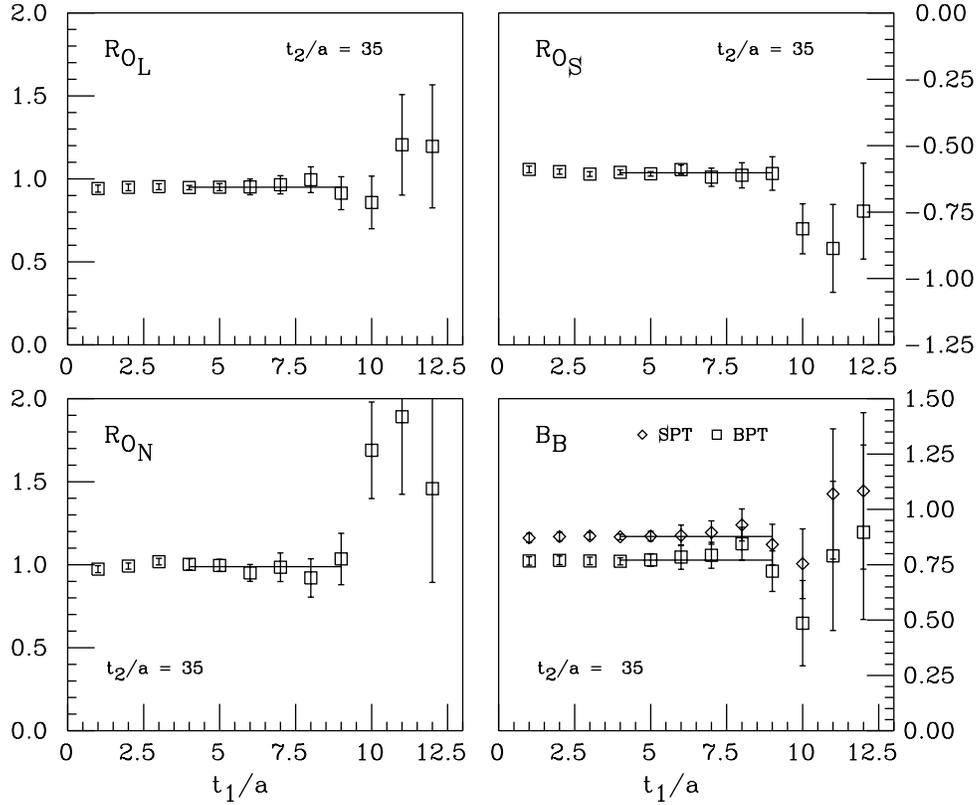}}
\end{picture}
\end{center}
\caption{\it{The ratios $R_{O_{L}}$, $R_{O_{S}}$, $R_{O_{N}}$ and 
$B_{B}$ from the combined-ratio method
are shown as a function of the time $t_1$, at 
$t_2/a= 35$.  They were  computed at $K=0.1432$, by using $A^D_\mu$,
eq.~(\protect\ref{sources}), 
 with $L_{s}=7$. The labels SPT and BPT 
refer to standard and boosted perturbation with the V-coupling
respectively.  The lines show the time-interval 
of the fits.}}
\label{fig:fourfig}
\end{figure}

In order to improve the isolation of the lightest meson state at short
time-distances, the ratios (\ref{ratio1}) and (\ref{ratio2}) using
single and double cubic smeared axial currents were
computed. The sizes of the cubes used
in our simulation are $L_{s}=5,7$ and $9$. From our previous studies,
$L_s=7$ and $9$ were shown to give a good isolation
of the lightest state  for    $t/a > 4-5$ or  $t/a < 36-35$ \cite{apeallton}.
For this reason, even though a plateau seems to set  in at much ealier times,
see fig. \ref{fig:fourfig}, 
the $B$-parameters were extracted
from  the ratios  (\ref{ratio1}) and (\ref{ratio2}) 
at fixed $t_2=34$--$36$ and  for $t_1 \ge 4$. 

In fig.~\ref{fig:fourfig},  in order to display the quality of our
data,  the results for 
$R_{O_{L}}$, $R_{O_{S}}$ and $R_{O_{N}}$ and $R_B=B_{B}$ as a function
of $t_1/a$, at $t_2$ fixed are shown in  a specific case.
With our large statistics, we are able to observe the
plateaux over large time-distances. Thus we were able to
fit the ratios  up  to $t_1/a =7$--$9$. 
On the basis of our previous experience,
this makes us confident that 
the lightest meson state has been isolated. 
\begin{table} \centering
\begin{tabular}{||c|c|c|c||}
\hline
\hline
\multicolumn{4}{||c||}{$B_{B}(m_{b})$}\\
\hline \hline
$K$&
\multicolumn{1}{c|}{$t_{2}=36$, $t_{1}=4$--$8$}&
\multicolumn{1}{c|}{$t_{2}=35$, $t_{1}=4$--$9$}&
\multicolumn{1}{c||}{$t_{2}=34$, $t_{1}=4$--$6$} \\
\hline \hline
0.1425& 0.796(15)& 0.774(19)& 0.79(3) \\
0.1432& 0.792(16)& 0.771(19)& 0.79(3) \\
0.1440& 0.787(18)& 0.77(2)& 0.79(3) \\ \hline
0.14543& 0.78(2)& 0.76(2)& 0.79(4)\\ \hline
0.14367& 0.79(2)& 0.77(2)& 0.79(3)\\
\hline\hline
\end{tabular}
\caption{\it{Values of $B_{B}(m_{b})$ obtained as explained in the text,
by using $A^D_\mu$, eq.~(\protect\ref{sources}), 
 with $L_{s}=7$. 
In order to obtain the renormalized operators, the boosted V-coupling
has been used.}}
\label{tab:bbtable}
\end{table}
\par
We have fitted the ratios (\ref{ratio1}) and (\ref{ratio2}) to a constant,
for several time-intervals.
As an example, in table \ref{tab:bbtable}
the  value of $B_{B}(m_{b})$,   at several values of
the light-quark Wilson parameter $K$, is given. 
Since the results are almost
independent of the value of the light quark mass,
they can be safely extrapolated.
The extrapolations to the chiral limit
($K_{c}=0.14543(1)$) and to the mass of the strange quark ($K_{s}=0.14367(6)$)
are  also presented in table \ref{tab:bbtable}.
\par
\begin{figure}
\vspace{9pt}
\begin{center}\setlength{\unitlength}{1mm}
\begin{picture}(160,80)
\put(30,-20){\epsfbox{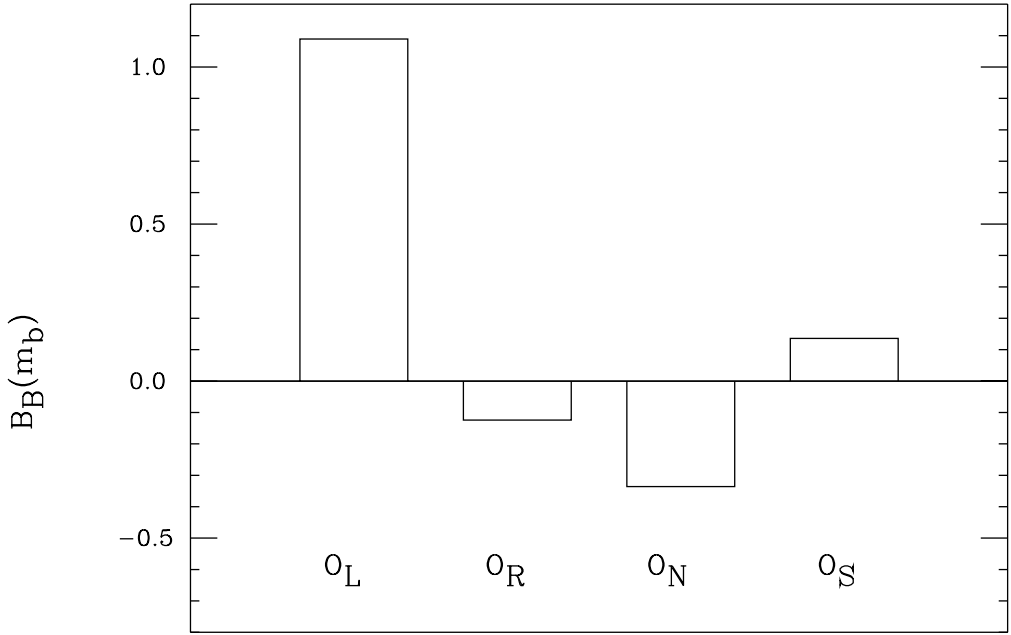}}
\end{picture}
\end{center}
\caption{\it{Comparison of the contributions of the operators 
$O_{L}$, $O_{R}$, $O_{N}$ and $O_{S}$ to the  $B$-parameter
$B_{B_d}(m_b)$.
The renormalized operator has been computed with the V-coupling,
using the option $M_1$.}}
\label{fig:histo}
\end{figure}

Finally, we  discuss the dependence of our results on the smearing size.
The value of all the matrix elements obtained with different
smearing sizes are very close. For example,
in the same  case as that considered in table \ref{tab:bbtable},
the values of  $B_{B}(m_b)$ in the chiral limit  
are $0.78(2)$, $0.76(2)$ and $0.74(3)$ for 
$L_{s}=5,7,9$,  respectively.
Within the statistical
errors,  a systematic shift of the
results with the smearing size cannot be appreciated. 
Thus we added  in quadrature 
the differences of the central values obtained with different
smearings  as an
error. In fig.~\ref{fig:histo}, in order to show the numerical 
importance of the different terms,  the contributions of 
the operators $O_{L,R,N,S}$ to the
$B$-parameter are shown. 
As expected, the operator $O_{L}$ gives the largest
contribution; the correction given by the three other
operators is of the order of 20\% and cannot be neglected.
\par
We now give the  final results by considering  first the ratio method. 
For the different operators the results are
\begin{eqnarray}
B_{O_L} = B_{O_R}= (0.94 \, \pm 0.02\, \pm 0.04)\ , \;\;\;\;\;\;\;\;\;\;
B_{O_N}  = (0.98 \, \pm 0.05\, \pm 0.04)\ ,\nonumber\\
\;\;\;\;\;\;\;\;\;\;
B_{O_S}  = -(0.60 \, \pm 0.01\, \pm 0.03)\ ,
\end{eqnarray}
where, as usual, the first error is the statistical one and the second is the
systematic one, which takes into account of differences coming from 
different choices of the interval of the fit  and of the smearing size. 
\par By using  in eq.~(\ref{zpt1}) 
the numerical values of the renormalization constants from
table~\ref{tab:zoi},  the values of $B_{B_d}$  given
in table~\ref{tab:bbb} were obtained. 
The results from  the ratio or the combined-ratio
 methods are indistinguishable. Notice, however, the large differences coming
from different choices of (equivalent) sets of perturbative renormalization
constants. From the results of the table, by using eq.~(\ref{bbhat}), 
it is straightforward to derive the renormalization
group invariant $\hat B_{B_d}$.
\begin{table} \centering 
\begin{tabular}{||c|c|c|c||}
\hline
\hline
\multicolumn{2}{||c|}{Options}&\multicolumn{1}{c|}{Ratio method}&
\multicolumn{1}{c||}{Combined-ratio method} \\
\hline
\hline
$\alpha_s^L$  Standard & $M_1$ &0.87(4)&0.87(5) \\ \cline{2-4}
 0.0796& $M_2$ &0.81(4) & 0.81(5)\\ \hline
$\alpha_s^L$  Boosted $u_0$ & $M_1$ &0.82(5)&0.81(5) \\ \cline{2-4}
 0.1458& $M_2$ &0.67(5)&0.67(4) \\ \hline
$\alpha_s^L$  Boosted V & $M_1$ &0.77(6)&0.76(5) \\ \cline{2-4}
 0.1800& $M_2$ &0.55(6)&0.54(4)\\ \hline
\hline
\hline
\end{tabular}
\caption{\it{Values of $B_{B_d}(m_b)$  for different choices of the
lattice coupling constants and options, see the text.}}
\label{tab:bbb}
\end{table}
\par
In table~\ref{tab:bbcomp},
our results for $B_{B}$  are compared to other determinations. 
In the table
$\hat{B}^{(1)}_{B_{d}}$ and $\hat{B}^{(2)}_{B_{d}}$ refers to the
renormalization group invariant $B$-parameters computed from
${B}_{B_{d}}(m_b)$ using the leading or next-to-leading formulae
respectively.  The leading formula is obtained
from eq.~(\ref{bbhat}) by dropping the term proportional
to $J_5$ and by using the one-loop expression for $\alpha_s$ (at one loop
$\alpha_s(m_b^2)=0.2342$). As for the results of the other groups,
irrespectively of the fact that they used leading or next-to-leading 
formulae to compute $B_{B_d}(m_b)$, we evolved  their value
from the scale used in the original paper to $m_b=5$ GeV using the
one loop evolution equations. For this reason some of the numbers
in the table may differ from those quoted in the corresponding references.
\par  
Our results are in good  agreement with those of
UKQCD  \cite{ukqcdbb},
when a similar recipe for the renormalization constants is
used~\footnote{ We recall that there are some
differences because they have ommitted the NLO corrections
and used by mistake the constants of ref.~\cite{borre1}
instead of ref.~\cite{borre2}. These differences are, however,
numerically unimportant.}. The UKQCD number reported
in table~\ref{tab:bbcomp} should be compared with the 
value obtained with the $M_2$ recipe and the $u_0$ coupling
constant, see table~\ref{tab:bbb}. 
With the exception of the result of ref.~\cite{christe},
which was obtained using the Wilson action for light quarks
and is surprisingly large, the static value
  is smaller than other
determinations of the same quantity obtained in the ``full" theory, i.e.
by extrapolating in the heavy quark mass $m_Q$ results obtained
for $m_Q \le m_b$ \cite{elc}--\cite{jlqcd}. 
A large value was also obtained by the  QCD sum rule calculation
of ref.~\cite{narison}.
\begin{table} \centering
\begin{tabular}{||c|c|c|c|c|c||}
\hline
\hline
$\hat{B}^{(1)}_{B_{d}}$&$\hat{B}^{(2)}_{B_{d}}$&
$B_{B_{d}}(m_{b})$&Authors&Ref.&Remarks\\
\hline 
\hline
1.45(22)&1.58(24)  & 0.99(15) 
                   &Narison {\it et al.}     & \cite{narison}& QSSR \\
1.29(7)& 1.40(7)  & 0.88(5)  &ELC                      & \cite{elc}    & Extrapolated Wilson $\beta=6.4$\\
1.30(9)&1.42(10)   & 0.89(7)  
                   &Soni {\it et al.}        & \cite{soni}   & Extrapolated Wilson $\beta=5.7-6.3$\\
1.31(7)&1.42(7)& 0.895(47)&JLQCD                    & \cite{jlqcd}  & Extrapolated Wilson $\beta=6.1$ \\
1.23(9)&1.34(10)& 0.840(60)&JLQCD                    & \cite{jlqcd}  & Extrapolated Wilson $\beta=6.3$ \\
1.01(6)&1.10(6)   & 0.69(4)  &UKQCD                    & \cite{ukqcdbb}  & Static      Clover      $\beta=6.2$ \\
1.42(6)&1.54(6) & 0.97(4)  &Christensen {\it et al.} & \cite{christe}& Static      Wilson    $\beta=6.0$\\
1.11(7)&1.21(8) & 0.76(5)  
&APE                      & This work     & Static      Clover  $\beta=6.0$ 
$M_1$\\
0.79(6)&0.86(6) & 0.54(4)  
&APE                      & This work     & Static      Clover  $\beta=6.0$ 
$M_2$\\
\hline\hline
\end{tabular}
\caption{\it{Values of the $B$-parameter as determined by previous
studies, and from the present one, are presented for comparison.
$\hat{B}^{(1)}_{B_{d}}$ and $\hat{B}^{(2)}_{B_{d}}$ refers to the
renormalization group invariant $B$-parameters computed from
${B}_{B_{d}}(m_b)$ using the leading or next-to-leading formulae
respectively. The results of this work are those obtained by using
the ``Boosted V" $\alpha_s^L$.}}
\label{tab:bbcomp}
\end{table}
\par
>From the study of the $B$-parameter as a function
of the light-quark mass, we also get 
\be
\mfrac{B_{B_{s}}}{B_{B_{d}}}\,=\,\mfrac{\hat{B}_{B_{s}}}{\hat{B}_{B_{d}}}\,=\,
1.01\, \pm\, 0.01
\ . \label{ratb} \ee
Notice that this
ratio is almost independent of choice of the different possible
options
 in the calculation of  the renormalization constants, see  sec.~\ref{bbdef}.
\par
Using the same set of gauge field configurations, the
following ratio has also been measured (see ref.~\cite{apeallton} for  details)
\be
\label{ratfb}
\mfrac{f_{B_{s}}}{f_{B_{d}}}\,=\,1.17\, \pm\, 0.03
\ee
Combining eqs.~(\ref{ratb}) and (\ref{ratfb}), one gets 
\be
\label{ratfbb}
\mfrac{f^{2}_{B_{s}}\, B_{B_{s}}}{f^{2}_{B_{d}}\, B_{B_{d}}}\,=\,
1.38\, \pm\, 0.07
\ . \ee
By using  eq.~(\ref{buri}) and the above result,
one obtains 
\be
\mfrac{x_{s}}{x_{d}}\,=\, (1.45\, \pm\, 0.13)\, 
\mfrac{\vert V_{ts}\vert^{2}}{\vert V_{td}\vert^{2}}
\ ,\label{qqq}
\ee
where, for the masses and lifetimes of
the $B$-mesons,  we have used the  values given in ref.~\cite{pdg}.
\par
\section{Conclusions} \label{conclu}

In this work we have computed  
the $B$-parameter of the $B$-meson in the static limit,
 on 600 gauge field configurations  with an improved action.
Given the large set of configurations, our 
statistical errors are rather small. The large statistics
 allows also a reduction of  the
systematic errors coming from the imperfect isolation of the
ground state. Errors coming from the extrapolation 
to the chiral limit are also
quite small.
Moreover, for this quantity, the error coming from the quenched
approximation was estimated to be negligible \cite{chiral}. 
Thus  the most important source of 
systematic error in our results is the determination of 
the renormalization constants, which are known so far only
in  one-loop lattice perturbation theory.
In order to  improve the accuracy of the lattice predictions   
for the $B$ parameter in the static theory, 
a non-perturbative computation of these constants 
is clearly necessary.

\section*{Acknowlegments}
We thank J.~Flynn and C.T.~Sachrajda for very useful discussions.
We acknowledge the partial support by the EC contract CHRX-CT92-0051.
V.G. acknowledges the partial support by CICYT under grant number AEN-96/1718
and the Phys. Dept. of the University ``La Sapienza" for hospitality.
G.M. acknowledges the partial support by M.U.R.S.T.

\end{document}